\documentclass[showpacs,twocolumn,preprintnumbers,amsmath,amssymb,pra]{revtex4}

\usepackage{graphicx}
\usepackage{dcolumn}
\usepackage{bm}

\renewcommand{\vec}[1]{{\mathbf #1}}
\newcommand{\vecgreek}[1]{{\mbox{\boldmath$ #1$}}}

\begin{document}

\title{Geometric Stability Spectra of Dipolar Bose Gases in Tunable Optical Lattices}

\author{John P. Corson$^1$}
\author{Ryan M. Wilson$^2$}
\author{John L. Bohn$^1$}

\affiliation{$^1$JILA, NIST and Department of Physics, University of Colorado, Boulder, Colorado 80309-0440, USA}
\affiliation{$^2$JQI, NIST and Department of Physics, University of Maryland, Gaithersburg, Maryland 20899-8410, USA}

\date{\today}

\begin{abstract}
We examine the stability of quasi-two-dimensional dipolar Bose-Einstein condensates in the presence of weak optical lattices of various geometries. We find that when the condensate possesses a roton-maxon quasiparticle dispersion, the conditions for stability exhibit a strong dependence both on the lattice geometry and the polarization tilt. This results in rich structures in the system's stability diagram akin to spectroscopic signatures. 
We show how these structures originate from the mode matching of rotons to the perturbing lattice. In the case of a one-dimensional lattice, some of the features emerge only when the polarization axis is tilted into the plane of the condensate. Our results suggest that the stability diagram may be used as a novel means to spectroscopically measure rotons in dipolar condensates.
\end{abstract}

\maketitle

\section{Introduction} \label{sec:Introduction}
The field of ultracold atoms has become a useful proving ground for theories of condensed matter systems. Progress with Bose-Einstein condensation in particular, both experimental and theoretical, has increased our understanding of superfluidity and long-range order and their dependence on interactions \cite{Leggett2001}. 
Recent years have seen increasing attention given to condensed systems whose constituent bosons interact via the dipole-dipole force, which is long-range and anisotropic. Such dipolar Bose-Einstein condensates (dBECs) have been achieved with atomic $^{52}$Cr \cite{Griesmaier2005}, $^{168}$Er \cite{Aikawa2012}, and $^{164}$Dy \cite{Lu2011}, and researchers are making active progress toward the condensations of more strongly interacting polar molecules \cite{Deiglmayr2008,Ni2010,Ospelkaus2010} and Rydberg atoms \cite{Heidemann2008}. The atomic dBEC experiments have observed various consequences of the dipole-dipole interaction, including geometry-dependent stability \cite{Koch2008,Lu2011}, $d$-wave collapse \cite{Lahaye2008,Aikawa2012}, and deconfinement-induced collapse \cite{Billy2012}.

Dipolar condensates are predicted to exhibit a roton-maxon structure in their dispersion relations \cite{Santos2003,ODell2003,Matuszewski2010}, qualitatively similar to that observed in superfluid $^4$He half a century ago \cite{Henshaw1958,Palevsky1958,Yarnell1959,Henshaw1961}. In the case of dBECs, the depth and location of the roton minimum can be tuned by variations of the interaction strength, density, and trapping parameters. A system becomes dynamically unstable when the roton energy is tuned to zero, although quantum and thermal fluctuations may destroy the condensate for small, but finite, roton energies \cite{Fischer2006,Boudjemaa2012}. Various signatures of rotonization have been predicted recently, such as 
reduced critical superfluid velocity \cite{Santos2003}, 
 abrupt transitions in Faraday patterns \cite{Faraday}, roton-length-scale oscillations of two-body correlations \cite{Sykes2012}, roton confinement \cite{Santos2013}, and short-wavelength immiscibility phases \cite{Wilson2012}. To date, none of these effects have been observed in dBECs, although a variant of Bragg spectroscopy recently measured the roton of a nondipolar gas with cavity-mediated long-range interactions \cite{Mottl2012}. Bragg spectroscopy, as calculated in Ref. \cite{Blakie2012}, should therefore be feasible in dipolar condensates.

The recently proposed method of stability spectroscopy \cite{Corson2013} is an appealing alternative for roton measurement that produces multiple signatures, while exploiting the inherent instability and sensitivity of rotonized systems. The general idea is to probe the rotonized dBEC with a weak lattice of periodicity $\lambda_{\rm L}$ and depth $s$, and then straightforwardly map the stability as a function of these lattice parameters. For quasi-one-dimensional (q1D) dBECs, it was demonstrated that the critical lattice depth $s_c$, above which no stable condensate exists, depends strongly on the periodicity and exhibits local minima whenever $\lambda_{\mathrm{L}}$ equals either half the roton wavelength or a low roton subharmonic. The stability boundary $s_c\left(\lambda_{\mathrm{L}}\right)$ hence constitutes a spectroscopic measurement of the roton wavelength, and it is easily measured because the lattice depth and spacing can be adjusted by varying respectively the intensity and angle of crossed off-resonant laser beams \cite{Grimm2000}. We emphasize that the necessary stability measurement is effectively binary: the condensate is either stable or unstable.

The goal of the present article is to generalize and extend the method of stability spectroscopy to quasi-two-dimensional (q2D) dBECs, elucidating the crucial roles played by lattice geometry and polarization tilt. We will show that, in a $1$D lattice, the central feature of the q1D case (at $\lambda_{\mathrm{L}}=\lambda_{\mathrm{rot}}$) persists in q2D, but the other features emerge only when the polarization axis is tilted into the trapping plane.
Additionally, we consider the stability spectra that result from $2$D triangular-lattice perturbations, where multiple roton signatures are most pronounced when the polarization is orthogonal to the trapping plane. Although our results are computed numerically, we explain the origins of the relevant stability structures by applying perturbation theory to the Gross-Pitaevskii and Bogoliubov de Gennes equations. Section \ref{sec: formalism} describes our mean-field formalism, Sec. \ref{sec: pert} outlines the perturbation theory employed in our analysis, Sec. \ref{sec: stability} presents and explains the stability spectra for several interesting q2D cases, and Sec. \ref{sec: conclusion} concludes our discussion.

\section{Mean-Field Formalism} \label{sec: formalism}

We consider a dilute gas of $N$ interacting bosons that is tightly confined in the $\hat z$ direction by a harmonic trap of frequency $\omega_t$ and moves freely in the $xy$ plane. The system is then subjected to a weak lattice perturbation in the $xy$ plane represented by $U(\vecgreek\rho)$.
We scale lengths and energies by the natural units of the problem, which are $\ell_t=\sqrt{\hbar/m\omega_t}$ and $\hbar \omega_t$ respectively. In the mean-field description of the system, the perturbed ground state $\Psi_0(\vec r)$ minimizes the grand canonical Hamiltonian
\begin{equation} \label{eq: 3d hamiltonian}
\begin{aligned}
H[\Psi] = & \int d^3r \Psi^*(\vec r)\left[-\frac{1}{2}\nabla^2+\frac{1}{2}z^2+U(\vecgreek\rho)-\mu_{3D} \right. \\& \quad \left.+\frac{1}{2}N\int d^3r'\left|\Psi(\vec r')\right|^2V_{3D}(\vec r - \vec r')   \right]\Psi(\vec r)
\end{aligned}
\end{equation}
where $\mu_{3D}$ is the chemical potential and $V_{3D}(\vec r)$ is the interaction pseudopotential for polarized dipoles \cite{Yi2000}
\begin{equation}
V_{3D}(\vec r)=4\pi a_s \delta\left(\vec r \right) +3a_{dd}\frac{1-\left(\hat d \cdot \hat r \right)^2}{r^3} .
\end{equation}
The interaction potential depends on the scattering length $a_s$ and the dipole length $a_{dd}=md^2/3\hbar^2$, as well as the polarization direction $\hat d$. The function $\Psi (\vec r)$ is assumed to have unit norm.

We make the simplifying assumption that both $U(\vecgreek\rho)$ and the interaction energy of the particles are small compared to the tight trapping energy. This allows for the single-mode approximation \cite{Petrov2000}, in which the order parameter is confined to zero-point oscillations along the $z$ direction. The $z$ dependence of the order parameter then factorizes out as $\Psi(\vec r)=\pi^{-\frac{1}{4}}\mathrm{e}^{-z^2/2}\psi(\vecgreek\rho)$. With this ansatz, Eq. \eqref{eq: 3d hamiltonian} reduces to
\begin{equation}\label{eq: 2d hamiltonian}
\begin{aligned}
H[\psi]=& \int d^2\rho \psi^*(\vecgreek \rho)\left[ -\frac{1}{2}\nabla^2_{\boldsymbol\rho}+U(\vecgreek \rho)+\frac{1}{2}-\mu_{3D}\right. \\& \left.
+\frac{1}{2}N\int d^2\rho '\left|\psi(\vecgreek \rho')\right|^2V(\vecgreek\rho -\vecgreek\rho')\right]\psi(\vecgreek\rho)
\end{aligned}
\end{equation}
where the q2D interaction $V$ is given in momentum space as
\begin{equation} \label{eq: VintK}
\begin{aligned}
\tilde V(\vec k)=\frac{4\pi}{\sqrt{2\pi}}&\left[  a_s+a_{dd}\left( \cos^2\alpha F_{\perp}\left(\frac{\vec k}{\sqrt{2}}\right)\right.\right. \\& \left.\left.\quad\quad\quad\quad\quad\quad
 +\sin^2\alpha F_{\parallel}\left(\frac{\vec k}{\sqrt{2}}\right) \right)\right]
\end{aligned}.
\end{equation}
The functions $F_{\perp}(\vec q)$ and $F_{\parallel}(\vec q)$ are defined by 
\begin{equation} \label{eq: F's}
\begin{aligned}
& F_{\perp}(\vec q)=2-3\sqrt{\pi}q\mathrm{e}^{q^2}\mathrm{erfc}(q) \\&
F_{\parallel}(\vec q)=-1+3\sqrt{\pi}\frac{\left(q_x^2\cos^2\eta+q_y^2\sin^2\eta\right)}{q}\mathrm{e}^{q^2}\mathrm{erfc}(q)
\end{aligned}
\end{equation}
with $\alpha$ and $\eta$ being respectively the polar and azimuthal angles defining the polarization direction $\hat d$ \cite{Ticknor2011}. Under the single-mode approximation, the ground state satisfies the q2D Gross-Pitaevskii equation
\begin{equation}\label{eq: GP}
\begin{aligned}
\mu \psi_0(\vecgreek \rho)=&-\frac{1}{2}\nabla^2_{\boldsymbol\rho}\psi_0(\vecgreek \rho) +U(\vecgreek\rho)\psi_0(\vecgreek\rho) \\&+N\int d^2\rho'\left|\psi_0(\vecgreek\rho')\right|^2V(\vecgreek\rho-\vecgreek\rho')\psi_0(\vecgreek\rho)
\end{aligned}
\end{equation}
with the effective q2D chemical potential $\mu\equiv \mu_{3D}-\frac{1}{2}$.

The dynamical excitations above the ground state $\psi_0$ are determined by solving the Bogoliubov de Gennes equations. These can be written compactly as
\begin{equation} \label{eq: BdG}
\begin{pmatrix}
H_0-\mu+C+X & X \\ -X & -H_0+\mu-C-X
\end{pmatrix}
\begin{pmatrix}
u_j \\ v_j
\end{pmatrix}
=E_j\begin{pmatrix}
u_j \\ v_j
\end{pmatrix}
\end{equation}
where $H_0=-\frac{1}{2}\nabla_{\boldsymbol\rho}^2+U(\vecgreek\rho)$ is the noninteracting single-particle Hamiltonian, $C[\chi](\vecgreek\rho)=N\int d^2\rho'\left|\psi_0(\vecgreek\rho')\right|^2V(\vecgreek\rho-\vecgreek\rho')\chi(\vecgreek\rho)$ describes direct interactions with the condensate, and $X[\chi](\vecgreek\rho)=N\int d^2\rho'\chi(\vecgreek\rho')\psi_0(\vecgreek\rho')V(\vecgreek\rho-\vecgreek\rho')\psi_0(\vecgreek\rho)$ is an integral operator describing exchange interactions. The functions $u_j$ and $v_j$ are subject to the usual normalization condition $\int d^2\rho\left(\left|u_j\right|^2-\left|v_j\right|^2 \right)=1$. The ground state $\psi_0$ satisfying Eq. \eqref{eq: GP} is dynamically unstable if one or more of the excitation energies $E_j$ is imaginary-valued, causing local collapse on a length scale set by the unstable mode \cite{Wilson2009}.

In the absence of the perturbation $U(\vecgreek\rho)$, the solutions to Eq.\eqref{eq: GP} and Eq. \eqref{eq: BdG} are well known. Translational invariance guarantees that momentum is a good quantum number. For simplicity, we will assume periodic boundary conditions over a rectangular domain of area $A$, which discretizes the momenta of the system. It is also convenient to introduce the integrated (over $z$) density $n_{2D}=N/A$. The unperturbed ground state and chemical potential are then $\psi_0^{(0)}(\vecgreek\rho)=1/\sqrt{A}$ and $\mu^{(0)}=\left.n_{2D}\tilde V(\vec k)\right|_{\vec k=0}$, respectively. The excitations are parameterized by their momentum quantum number $\vec k$, and are given by \cite{Fetter}
\begin{equation} \label{eq: BdG modes}
\begin{pmatrix}
u_{\vec k}^{(0)}(\vecgreek\rho) \\
v_{\vec k}^{(0)}(\vecgreek \rho)
\end{pmatrix} = 
\begin{pmatrix}
\sqrt{\frac{k^2/2+n_{2D}\tilde V(\vec k)}{2\omega(\vec k)}+\frac{1}{2}} \\
-\mathrm{sgn}(\tilde V(\vec k))\sqrt{\frac{k^2/2+n_{2D}\tilde V(\vec k)}{2\omega(\vec k)}-\frac{1}{2}}
\end{pmatrix}
\varphi_{\vec k}(\vecgreek\rho)
\end{equation}
where $\varphi_{\vec k}(\rho)\equiv \mathrm{e}^{i\vec k \cdot \boldsymbol\rho}/\sqrt{A}$ and  $E^{(0)}_{\vec k}=\omega(\vec k)\equiv \sqrt{k^2/2\left(k^2/2+2n_{2D}\tilde V(\vec k)\right)}$ is the Bogoliubov spectrum. 

For certain densities and interaction parameters, the dispersion relation $\omega(\vec k)$ may contain a local minimum at nonzero momentum. The corresponding mode is referred to as a roton mode. When the dipoles are polarized orthogonal to the plane of motion, the interaction and dispersion depend only on the momentum magnitude $k$, thereby causing the set of roton modes to form a ring of radius $k_{\rm rot}$ in $k$-space. Tilting the dipoles into the plane, however, makes both the interaction and disperion anisotropic. As an example, consider the case in which the polarization is tilted somewhat towards the $y$ axis. Modes that propagate along $\hat y$ create density antinodes along lines of constant $x$, thereby accumulating dipoles in the higher-energy side-to-side configuration. Conversely, modes that propagate along $\hat x$ tend to accumulate dipoles along lines of constant $y$, where the dipoles are somewhat head-to-tail because of the tilt. In a general sense, modes that propagate along the tilt projection tend to have higher energy than those that propagate in a direction orthogonal to the tilt projection, and there is a continuous transition between the two as the direction of a mode is varied. This anisotropy of dispersion has been shown to lead to anisotropic superfluidity \cite{Ticknor2011} and coherence \cite{Ticknor2012}, as well as striped immiscibility states in binary dipolar condensates \cite{Wilson2012}. We will see in Section \ref{sec: 1D latt} that this anisotropy leads to emergent features in the stability spectrum of a q2D dipolar condensate in a 1D lattice.

\section{Perturbation Theory} \label{sec: pert}

Reference \cite{Corson2013} demonstrates that the main features of a q1D stability spectrum can be understood in the context of simple perturbation theory. We have found that this is also true for q2D spectra. In this section, we briefly develop the essentials of perturbation theory which are necessary to understand the results of Sec. \ref{sec: stability}. We will establish first-order perturbative results for the order parameter and mean-field potential, and then write down the first-order perturbation equation for Bogoliubov modes. Further details regarding rigorous perturbation theories applied to the Gross-Pitaevskii and Bogoliubov de Gennes equations can be found in Refs. \cite{Wilson2008,Taylor2003,Gaul2011,Liang2008, Lugan2011}.

To study the response of the condensate to the lattice perturbation $U(\vecgreek\rho)$, we expand both the order parameter and chemical potential in perturbation series as $\psi=\psi^{(0)}+\psi^{(1)}+\dots$ and $\mu=\mu^{(0)}+\mu^{(1)}+\dots$, respectively. To first order, the Gross-Pitaevskii equation \eqref{eq: GP} reduces to the linear perturbation equation
\begin{equation} \label{eq: GP pert}
\begin{aligned}
\mu^{(1)}\psi^{(0)}=& -\frac{1}{2}\nabla_{\boldsymbol\rho}^2\psi^{(1)}(\vecgreek\rho)+U(\vecgreek\rho)\psi^{(0)}(\vecgreek\rho) \\& \quad\quad\quad
+2n_{1D}\int d^2\rho'\psi^{(1)}(\vecgreek\rho')V(\vecgreek\rho-\vecgreek\rho')
\end{aligned}
\end{equation}
under the requirement that the order parameter remains normalized to first order, implying that $\left\langle \psi^{(0)}| \psi^{(1)}\right\rangle=0$. Multiplying both sides of Eq. \eqref{eq: GP pert} by $\psi^{(0)}$ and integrating yields the correction to the chemical potential
\begin{equation}
\mu^{(1)}=\left\langle \psi^{(0)} | U | \psi^{(0)}\right\rangle .
\end{equation}
We compute the correction to the order parameter by expanding $\psi^{(1)}$ in the complete basis of complex exponential functions $\left\{\varphi_{\vec k}  \right\}$ introduced below Eq. \eqref{eq: BdG modes}. We then find that the expansion amplitudes are given by
\begin{equation} \label{eq: psi pert}
\left\langle \varphi_{\vec k}|\psi^{(1)}\right\rangle = -\frac{\left\langle\varphi_{\vec k}| U |\psi^{(0)}\right\rangle}{\varepsilon(\vec k)}
\end{equation}
where $\varepsilon(\vec k)\equiv k^2/2+2n_{1D}\tilde V(\vec k)$ is the Hartree-Fock energy \cite{Pitaevskii}. Given that $\omega(\vec k)=\sqrt{\varepsilon(\vec k)k^2/2}$, the presence of a roton mode in the Bogoliubov dispersion implies a local minimum in the Hartree-Fock spectrum at a comparable value of $\vec k$. For soft-roton systems, the locations of these minima very nearly coincide. It is easy to show that $\varepsilon(\vec k)\rightarrow 2\mu^{(0)}$ in the zero-momentum limit, indicating that Eq. \eqref{eq: psi pert} is well defined even for small $\vec k$ whenever the unperturbed condensate is stable.

For the purposes of this paper, it is useful to define the combined potential experienced by an atom at position $\vecgreek\rho$, which is the sum of the external and mean-field potentials. We evaluate this potential to first order from \eqref{eq: psi pert}, which results in
\begin{equation} \label{eq: Uc}
\begin{aligned}
U_c(\vecgreek\rho)& \equiv U(\vecgreek\rho)-\mu+N \int d^2\rho'\left|\psi(\vecgreek\rho')\right|^2V(\vecgreek\rho-\vecgreek\rho') \\&
= \sum_{\vec k \neq 0}\mathrm{e}^{i\vec k\cdot \boldsymbol\rho}\frac{k^2}{2\varepsilon(\vec k)}\left\langle \varphi_{\vec k} | U | \psi^{(0)}\right\rangle +\mathcal{O}\left( U^2 \right)
\end{aligned}
\end{equation}
after some algebra. A typical lattice potential projects onto only a few basis modes $\varphi_{\vec k}$, which greatly simplifies the formulae for both $\psi^{(1)}$ and $U_c$ in practice. In the absence of interactions, the combined potential is of course equal to the external potential itself. However, dipolar interactions may cause the system to rotonize, in which case the mean-field potential may amplify the perturbation by a factor of order $k^2/2\varepsilon(\vec k)$, the magnitude of which depends strongly on the modes that compose the lattice.

\begin{figure*}
\includegraphics[width=.9\textwidth]{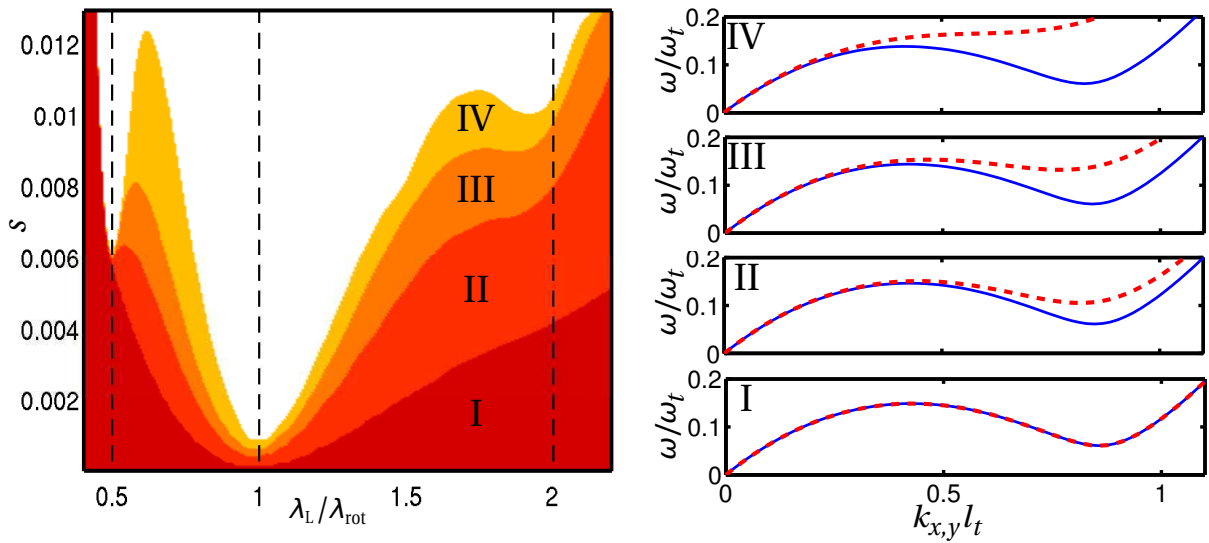}
\caption{(Color online) 
Stability spectra for q2D dBECs in a weak lattice with varied polarization tilts and scattering lengths, chosen in such a way that $\omega(k_x,0)$ is similar in each case. The shaded regions correspond to stable, condensed ground states. The experimental parameters are for $^{164}$Dy trapped at $\omega_t = 2\pi \times 10$ kHz with a density $n_{3D}\approx 10^{15}$ cm$^{-3}$ and dipole moment $d=10\mu_B$. Cases I-IV correspond respectively to polarization tilt $\alpha$ (towards $\hat y$) and scattering length $a_s$ values of 
$\{\alpha,a_s\}_{\mathrm{I}}=\{0^{\circ},-82a_0\}$, $\{\alpha,a_s\}_{\mathrm{II}}=\{7^{\circ},-79.7a_0\}$, $\{\alpha,a_s\}_{\mathrm{III}}=\{10^{\circ},-77.35a_0\}$, and $\{\alpha,a_s\}_{\mathrm{IV}}=\{15^{\circ},-71.6a_0\}$.
The right-hand panels depict the dispersion relations for these cases, with the blue (solid) lines corresponding to $\omega(k_x,0)$ and the red (dashed) lines corresponding to $\omega(0,k_y)$. For each case, the roton wavelength (for modes parallel to $\hat x$) is approximately $\lambda_{\mathrm{rot}}\approx 575$ nm. The vertical dashed lines on the stability spectrum denote $\lambda_{\mathrm{rot}}/2$, $\lambda_{\mathrm{rot}}$, and $2\lambda_{\mathrm{rot}}$, which approximately identify the spectral features.
}
\label{fig: stab tilts}
\end{figure*}

To study the response of quasiparticle energies to the perturbation, we now apply a simple perturbation theory to the Bogoliubov de Gennes equations. We expand $E_j$, $u_j$, and $v_j$ in perturbation series, and then substitute into Eq. \eqref{eq: BdG} to find the perturbation equation:
\begin{equation}\label{eq: BdG pert}
\begin{aligned}
&\begin{pmatrix}
\omega(\vec k)+\frac{1}{2}\nabla^2-X^{(0)} & -X^{(0)} \\
-X^{(0)} & -\omega(\vec k)+\frac{1}{2}\nabla^2-X^{(0)}
\end{pmatrix}
\begin{pmatrix}
u_{\vec k}^{(1)} \\ v_{\vec k}^{(1)}
\end{pmatrix} = \\&
\begin{pmatrix}
U_c^{(1)}+X^{(1)} & X^{(1)} \\
X^{(1)} & U_c^{(1)}+X^{(1)}
\end{pmatrix}
\begin{pmatrix}
u_{\vec k}^{(0)} \\ v_{\vec k}^{(0)}
\end{pmatrix} -
E_{\vec k}^{(1)}
\begin{pmatrix}
u_{\vec k}^{(0)} \\ -v_{\vec k}^{(0)}
\end{pmatrix}
\end{aligned}
\end{equation}
where $X^{(0)}[\chi](\vecgreek\rho)=N\int d^2\rho' \chi(\vecgreek\rho')\psi^{(0)2}V(\vecgreek\rho-\vecgreek\rho')$ describes exchange interactions with the unperturbed condensate, $U_c^{(1)}(\vecgreek\rho)$ is the first-order combined potential given by Eq. \eqref{eq: Uc}, and $X^{(1)}[\chi](\vecgreek\rho)=N\int d^2\rho'\chi(\vecgreek\rho')\psi^{(0)}V(\vecgreek\rho-\vecgreek\rho')\left(\psi^{(1)}(\vecgreek\rho')+\psi^{(1)}(\vecgreek\rho)\right)$ describes exchange interactions with the condensate perturbation. Our present interest lies only in the energy shift $E_{\vec k}^{(1)}$, which we can isolate by acting on both sides of Eq. \eqref{eq: BdG pert} by the operator $\int d^2\rho \left(u_{\vec k'}^{(0)},v_{\vec k'}^{(0)} \right)$ for any $\vec k'$ satisfying the degeneracy condition $\omega(\vec k')=\omega(\vec k)$. First-order energy shifts are then determined by the Hermitian matrix on the right-hand side of Eq. \eqref{eq: BdG pert}, which we denote compactly as 
\begin{equation} \label{eq: A matrix}
\mathcal{A}\equiv \begin{pmatrix}
U_c^{(1)}+X^{(1)} & X^{(1)} \\
X^{(1)} & U_c^{(1)}+X^{(1)}
\end{pmatrix} .
\end{equation}
Our analysis will focus primarily on the softening of roton modes, which generally are closely connected to system instability.

\section{Stability Spectra} \label{sec: stability}

\subsection{One-Dimensional Lattice} \label{sec: 1D latt}

We now consider the stability spectrum of a rotonized q2D dipolar condensate that is perturbed by a one-dimensional lattice of tunable depth and lattice spacing. Without loss of generality, we focus on the special case in which the lattice is directed along the $x$-axis:
\begin{equation} \label{eq: 1D latt}
U(\vecgreek\rho)=s \cos\left(k_{\mathrm{L}} x\right) .
\end{equation}
Such a potential may be generated by a pair of crossed off-resonant beams of wavelength $\lambda_{\mathrm{Las}}$ and angle $\theta$, with the lattice spacing determined by the relation $\lambda_{\mathrm{L}}=2\pi/k_{\mathrm{L}}=\lambda_{\mathrm{Las}}/2\sin(\theta/2)$. The depth parameter is proportional to the single-beam intensity $I_0$  via $s = -\mathrm{Re}\left\{\alpha(\omega)\right\}I_0/4\epsilon_0c\hbar\omega_t$, where $\mathrm{Re}\left\{\alpha(\omega)\right\}$ is the atomic polarizability. The depth and spacing of the lattice may be tuned by respectively varying the intensity $I_0$ and angle $\theta$ \cite{Grimm2000}. For each spacing $\lambda_{\mathrm{L}}$, the dBEC destabilizes at depths above a certain critical value $s_{\mathrm{crit}}(\lambda_{\mathrm{L}})$.

Our numerical method maps the stability of a system by exploiting two intrinsic symmetries of the 1D lattice \eqref{eq: 1D latt}. First of all, the system remains invariant with respect to translations in $y$. The order parameter of a stable perturbed ground state should thus be independent of this variable, although a collapsed state will generally exhibit a spontaneous breaking of this symmetry due to local collapse \cite{Wilson2009}. With this in mind, we numerically solve the Gross-Pitaevskii equation \eqref{eq: GP} using the conjugate gradients method \cite{Press} under the assumption of translational invariance along $y$. We then use the computed ground state to numerically solve the Bogoliubov de Gennes equations \eqref{eq: BdG}, which themselves are block-diagonal due to both parity in $x$ and translational invariance in $y$. The signature of a collapsed state is the existence of imaginary-valued Bogoliubov excitation energies for an assumed $y$-independent order parameter, or alternatively local collapse. Our numerical grid consists of $2^9$ points along each of $x,y\in[-24\lambda_{\mathrm{rot}},24\lambda_{\mathrm{rot}}]$.

Figure \ref{fig: stab tilts} shows the numerically-computed stability spectra for several different rotonized dBECs. The different stability boundaries correspond to different polarization tilts towards the $y$-axis ($\eta=90^{\circ}$, in Eq. \eqref{eq: F's}), whose dispersions $\omega(k_x,0)$ and $\omega(0,k_y)$ are depicted in the subplots. For each case, we have chosen the tilt $\alpha$ and scattering length $a_s$ in such a way that the dispersions along $\hat x$, most particularly the depth of the roton mode, are comparable in each case as shown by the solid blue lines. This allows us to isolate the consequences of polarization tilt from those related to roton depth. As described earlier, polarization tilt causes the modes directed along the tilt projection ($\hat y$ in this case) to have higher energy than modes that are orthogonal to the tilt projection (along $\hat x$). 

In Fig. \ref{fig: stab tilts}, we observe that the central stability feature at $\lambda_{\mathrm{L}}\approx\lambda_{\mathrm{rot}}$ appears to exist independent of polarization tilt. It originates in the strong amplification of the mean-field potential, as occurs in q1D systems \cite{Corson2013}. This may be seen by evaluating the combined potential \eqref{eq: Uc} given the 1D lattice \eqref{eq: 1D latt}:
\begin{equation} \label{eq: Uc 1d}
U_c(\vecgreek\rho)=s\left(\frac{k_{\mathrm{L}}^2}{2\varepsilon(k_{\mathrm{L}}\hat x)}\right)\cos(k_{\mathrm{L}}x) +\mathcal{O}\left(s^2\right) .
\end{equation}
In each case considered, the Hartree-Fock energy $\varepsilon(\vec k)$ has a shallow local minimum at $k_{\mathrm{L}}\hat x \approx k_{\mathrm{rot}}\hat x$ due to rotonization. Equation \eqref{eq: Uc 1d} thus indicates that the mean-field strongly amplifies the perturbation by a factor of $k_{\mathrm{rot}}^2/2\varepsilon(k_{\mathrm{rot}}\hat x)\gg 1$ for $\lambda_{\mathrm{L}}\approx \lambda_{\mathrm{rot}}$. This mean-field enhancement of the lattice probe is responsible for the central stability dip in each case. A similar effect was recently observed in simulations of nonrotonized dBEC superflow in a weak lattice \cite{Kuhn2013}. In that context, mean-field enhancement originates from the polarization tilt (towards $\hat y$) effectively lowering the trapped-gas analogue of $\varepsilon ( k_{\mathrm{L}}\hat x)$ for fixed scattering length and lattice spacing. 

The appearance of stability dips at $\lambda_{\mathrm{L}} = \lambda_{\mathrm{rot}}/2$ and $\lambda_{\mathrm{L}}\approx 2\lambda_{\mathrm{rot}}$ depends strongly on the polarization tilt angle $\alpha$. This can be understood in terms of the mode-matching of Bogoliubov roton modes by the Hermitian matrix $\mathcal{A}$ defined in Eq. \eqref{eq: A matrix}. Both $\left\langle \varphi_{\vec k'}| U_c^{(1)} | \varphi_{\vec k}\right\rangle$ and $\left\langle \varphi_{\vec k'}| X^{(1)}| \varphi_{\vec k}\right\rangle$ are proportional to $\delta_{k_y',k_y}\delta_{\left|k_x'-k_x \right|,k_{\mathrm{L}}}$, which defines the mode-matching conditions of the perturbation. Degenerate modes satisfying these conditions experience a first-order energy shift. Given the reflection symmetry of the perturbation \eqref{eq: 1D latt}, a mode-matched pair will separate into even and odd solutions in $x$. For $s>0$, the odd solution will lower in energy because it accumulates atoms in the minima of the combined potential \eqref{eq: Uc 1d}, as illustrated in Ref. \cite{Corson2013}; conversely, the even solution will increase in energy. This mixing and splitting of degenerate, matched modes is similar to that of ``staggered modes'' which appear at the Brillouin zone edge in single-particle band theory \cite{Morsch2006}, although we are considering the dispersion of quasiparticles.

\begin{figure}
\includegraphics[width=0.3\textwidth]{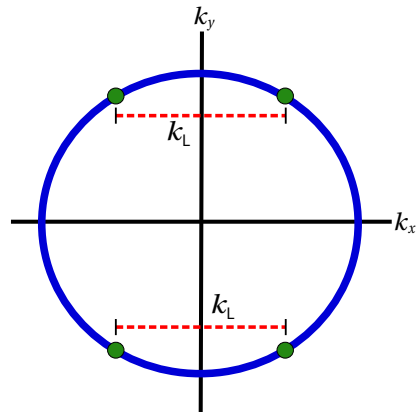}
\caption{(Color online) Depiction of roton modes for the case of zero polarization tilt ($\alpha =0$). Degenerate modes satisfying $k_y=k_y'$ and $\left|k_x-k_x'\right|=k_{\mathrm{L}}$ (indicated by pairs of green dots) experience first-order energy shifts.}
\label{fig: mode match}
\end{figure}

Because roton modes are local minima of the system dispersion, we expect hastened destabilization for all lattice periodicities in which low-energy roton modes soften to first order. Figure \ref{fig: mode match} depicts the $k$-space ring of degenerate rotons for the $\alpha=0$ case. For $k_{\mathrm{L}}\in \left(0,2k_{\mathrm{rot}}\right)$, there are always two pairs of rotons,
\[ \left(\pm k_{\mathrm{L}}/2,\sqrt{k_{\mathrm{rot}}^2-\left(  k_{\mathrm{L}}/2\right)^2}\right) 
\]
and
\[
\left(\pm k_{\mathrm{L}}/2,-\sqrt{k^2_{\mathrm{rot}}-\left(k_{\mathrm{L}}/2\right)^2}   \right) ,
\]
which mode match. For $k_{\mathrm{L}}=2k_{\mathrm{rot}}$, these two pairs coalesce into a single pair at $\vec k=(\pm k_{\mathrm{rot}},0)$. The fact that there is always a low-energy roton pair to soften for $\lambda_{\mathrm{L}}\geq \lambda_{\mathrm{rot}}/2$ implies that the BEC is fairly unstable to collapse at small $s$ for all applied lattice wavelengths in this range. The stability boundary I in Fig. \ref{fig: stab tilts} is therefore low and relatively featureless except for the central dip already accounted for.
By contrast, tilting the polarization towards $\hat y$ raises the relative energies of rotons with a nonzero $k_y$ component. With increasing tilt $\alpha$ (and fixed $\omega(k_{\mathrm{rot}},0)$), we then expect an emergent stability dip when the \emph{lowest}-energy rotons (along $\pm\hat x$) mode-match. This occurs when $k_{\mathrm{L}}=2 k_{\mathrm{rot}}$ ($\lambda_{\mathrm{L}}=\lambda_{\mathrm{rot}}/2$), explaining the other prominent feature in Fig. \ref{fig: stab tilts} case IV. This is in direct analogy with the corresponding stability dip in the q1D scenario \cite{Corson2013}, in which case excitations with a $\hat y$ component have higher relative energies due to tight trapping, rather than polarization tilt. We note that we have observed the emergence of the $\lambda_{\mathrm{rot}}/2$ stability feature in fully-$3$D simulations of flat pancake dBECs, where $\lambda_{\mathrm{rot}}$ is less well defined. Thus, the emergent stability dip is observable in the experimentally-realistic scenario of a rotonized trapped gas.

The smaller feature at $\lambda_{\mathrm{L}}\approx 2\lambda_{\mathrm{rot}}$ in Fig. \ref{fig: stab tilts} derives from the interplay of second-order effects, such as (modest) mean-field amplification and the mode-matching of degenerate rotons and phonons along $\hat x$. The shift to a slightly shorter wavelength is a consequence of the Hartree-Fock minimum occuring at a somewhat larger momentum than $k_{\mathrm{rot}}$. Moreover, the first mode mixture to destabilize in this case may not even include the precise roton minimum, since the (degenerate) second-order energy shifts responsible for destabilization depend on the dispersion as a whole. This complicates the precise analytical determination of this feature's location beyond the approximation $\lambda_{\mathrm{L}}\approx 2\lambda_{\mathrm{rot}}$; however, this approximation improves as the unperturbed roton mode softens in energy.

The stability structures discussed in this section tend to disappear when the polarization is tilted towards the lattice vector (along $\hat x$). Such a tilt increases the Hartree-Fock energy $\varepsilon \left(k_{\mathrm{L}}\hat x\right)$, thereby reducing the effect of mean-field amplification in Eq. \eqref{eq: Uc 1d}. Moreover, the lowest-energy rotons (now along $\pm \hat y$) only mode-match to each other when $k_{\mathrm{L}}$ is vanishingly small. As a result, the stability boundary of such cases tends to be relatively structureless and is not depicted here.

\subsection{Triangular Lattice}

\begin{figure}
\includegraphics[width=0.3\textwidth]{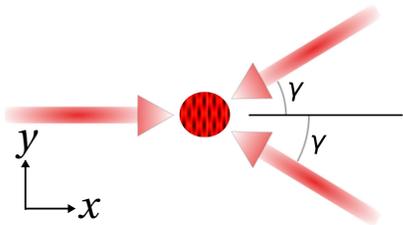}
\caption{(Color online)
Schematic for the alignment of lasers (in the $xy$ plane) used to produce the triangular lattice that we consider.
}
\label{fig: tri latt}
\end{figure}

For the case of a $1$D lattice imposed on a rotonized q2D dBEC, a polarization tilt is required for the appearance of multiple stability dips. We now demonstrate that a triangular lattice perturbation causes multiple stability structures to emerge even in the $\alpha=0$ case.
One may produce a triangular lattice using three crossed, off-resonant beams as depicted in Fig. \ref{fig: tri latt}. Without loss of generality, we assume a stationary beam directed along $\hat x$, with symmetrically-placed beams making an incidence angle $\gamma$ with the $x$-axis as shown \cite{symmetry}. Assuming equal polarization and intensity among the beams, the corresponding potential is \cite{Coordinates}
\begin{equation} \label{eq: tri latt}
\begin{aligned}
U(\vecgreek\rho)=s & \left[2\cos\left(k_{\mathrm{Las}}(1+\cos\gamma)x\right)\cos\left(k_{\mathrm{Las}}(\sin\gamma)y\right)\right.\\& \left.
+\cos(2k_{\mathrm{Las}}\left(\sin\gamma\right) y)\right]
\end{aligned}
\end{equation}
where $k_{\mathrm{Las}}=2\pi/\lambda_{\mathrm{Las}}$ is the laser wave number and $s$ is the AC Stark shift factor defined below Eq. \eqref{eq: 1D latt} \cite{Pethick}.  As the notion of lattice spacing is ambiguous for this potential, we instead vary the crossing angle $\gamma\in\left(0 , 180\right)^{\circ}$ and identify the stability boundary $s_{\mathrm{crit}}(\gamma)$.

Due to the lack of continuous translational invariance of the triangular lattice \eqref{eq: tri latt}, the block-diagonals of the Bogoluibov de Gennes eigenvalue problem \eqref{eq: BdG} are large compared to those of the $1$D lattice scenario in Sec. \ref{sec: 1D latt}. This makes repeated diagonalization more computationally expensive. We instead assess stability directly from the Gross-Pitaevskii energy functional \eqref{eq: 2d hamiltonian}. Implementing a conjugate gradients algorithm \cite{Press}, we compute the energy minimum (converged to a relative error of $10^{-12}$, if a minimum exists) and use local collapse as the signature of instability. Our simulations use a grid of $2^8$ points along $x\in [-16\lambda_{\mathrm{rot}},16\lambda_{\mathrm{rot}}]$ and $2^7$ points along $y\in [-8\lambda_{\mathrm{rot}},8\lambda_{\mathrm{rot}}]$. We choose a fixed, horizontal laser mode $k_{\mathrm{Las}}$ from our $k$-space grid, and compute stability only for angles $\gamma$ associated with $k$-grid modes satisfying $\left|\left|\vec k\right|-k_{\mathrm{Las}}  \right|<dk_x/2$, where $dk_x=\pi/16\lambda_{\mathrm{rot}}$ is the numerical grid spacing of $k_x$ modes. This enforces in an approximate way our assumption that the three lasers forming the lattice have equivalent wavelengths, while guaranteeing that the lattice satisfies the assumed periodic boundary conditions of our grid. We further employ shape-preserving piecewise cubic interpolation to the smoothly-varying stability boundary for angles that cannot be produced on our finite, periodic spatial grid.  We find that these numerical approximations are sufficient to resolve the important spectral features of the stability diagram. 

\begin{figure}
\includegraphics[width=0.45\textwidth]{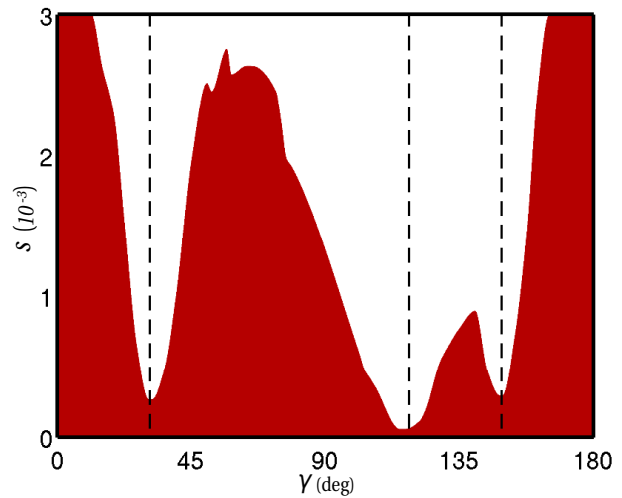}
\caption{(Color online) Stability Spectrum for a (zero-polarization-tilt) q2D dBEC subjected to a triangular lattice perturbation. The shaded region corresponds to a stable, condensed ground state. The experimental parameters are identical to those of case I in Fig. \ref{fig: stab tilts}. The assumed laser wavelength is $\lambda_{\mathrm{Las}}=595$ nm. The vertical dotted lines denote the predicted locations of stability dips, based on first-order perturbation theory applied to the Gross Pitaevskii equation. We have employed piecewise cubic interpolation to smooth the stability boundary. }
\label{fig: stab tri}
\end{figure}

\begin{figure}
\includegraphics[width=0.45\textwidth]{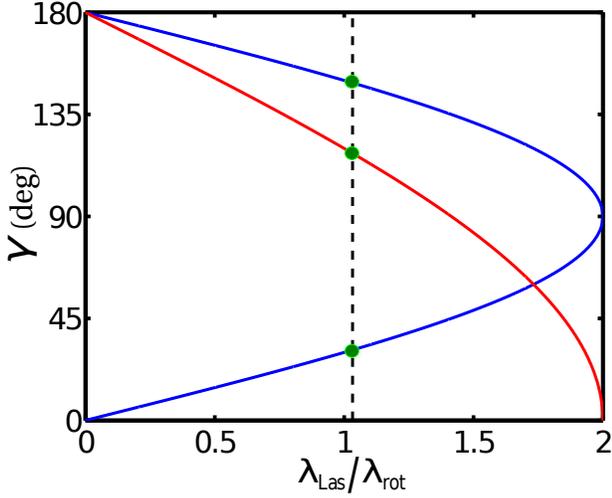}
\caption{(Color online) Predicted locations of stability structures as a function of laser wavelength $\lambda_{\mathrm{Las}}$. The symmetric blue lines are solutions to $\sin\gamma=\lambda_{\mathrm{Las}}/2\lambda_{\mathrm{rot}}$, and the red line is the solution to $\cos\frac{\gamma}{2}=\lambda_{\mathrm{Las}}/2\lambda_{\mathrm{rot}}$. Angles corresponding to the laser wavelength used for Fig. \ref{fig: stab tri} are indicated by green dots, and they define the locations of the vertical lines overlain on the triangle-lattice stability spectrum in Fig. \ref{fig: stab tri}.}
\label{fig: latt locations}
\end{figure}

Figure \ref{fig: stab tri} shows the triangle-lattice stability spectrum of the same $\alpha=0$ dBEC as was used in Fig. \ref{fig: stab tilts}. We observe three main features, each of which originates entirely from mean-field amplification. We compute the first-order combined potential by substituting Eq. \eqref{eq: tri latt} into Eq. \eqref{eq: Uc}:
\begin{equation} \label{eq: Uc tri}
\begin{aligned}
U_c(\vecgreek\rho)=s&\left[2 A_{k_{\mathrm{Las}},\gamma}\cos\left(k_{\mathrm{Las}}(1+\cos\gamma)x\right)\cos\left(k_{\mathrm{Las}}(\sin\gamma) y  \right)\right. \\&
\left.+B_{k_{\mathrm{Las}},\gamma}\cos\left(2k_{\mathrm{Las}}(\sin\gamma)y\right)   \right] +\mathcal{O}\left(s^2\right)
\end{aligned}
\end{equation}
where
\begin{equation} \label{eq: A and B}
\begin{aligned}
 & A_{k_{\mathrm{Las}},\gamma}\equiv \frac{\left(2k_{\mathrm{Las}}\cos(\frac{\gamma}{2})\right)^2}{2\varepsilon\left(2k_{\mathrm{Las}}\cos(\frac{\gamma}{2})\right)}\\&
B_{k_{\mathrm{Las}},\gamma}\equiv \frac{\left(2k_{\mathrm{Las}}\sin\gamma\right)^2}{2\varepsilon(2k_{\mathrm{Las}}\sin\gamma)}
\end{aligned}.
\end{equation}
Given that $\varepsilon (k)$ has a shallow local minimum at $k \approx k_{\mathrm{rot}}$, a component of the perturbation \eqref{eq: tri latt} is strongly amplified when either $\sin\gamma \approx k_{\mathrm{rot}}/2k_{\mathrm{Las}}$ or $\cos(\frac{\gamma}{2})\approx k_{\mathrm{rot}}/2k_{\mathrm{Las}}$ is satisfied. Figure \ref{fig: latt locations} plots the three solutions to these equations on the domain $\gamma\in (0,180)^{\circ}$ as functions of $\lambda_{\mathrm{Las}}/\lambda_{\mathrm{rot}}=k_{\mathrm{rot}}/k_{\mathrm{Las}}$, which exist only for $\lambda_{\mathrm{Las}}\leq 2\lambda_{\mathrm{rot}}$ (or, equivalently, $k_{\mathrm{Las}}\geq k_{\mathrm{rot}}/2$). 
For the particular laser wavelength relevant to Fig. \ref{fig: stab tri} ($\lambda_{\mathrm{Las}}/\lambda_{\mathrm{rot}}=1.03$), three solutions are identified (green dots) which determine the values of $\gamma$ where stability minima should occur. These predicted minima are depicted in Fig. \ref{fig: stab tri} with vertical lines.
As we can see, the observed locations of the stability dips agree quite well with the prediction. The stability boundary near $\gamma\approx 120^{\circ}$ is somewhat lower because it originates from the mean-field enhancement of the first term in Eq. \eqref{eq: Uc tri}, which has an additional factor of $2$.

It is interesting that these locations can be tuned by varying the laser wavelength, with the constraint that the wavelength remain detuned from internal atomic transitions. We note that the two structures which are symmetric with respect to the angle $90^{\circ}$ (located at $\gamma_s\leq 90^{\circ}$ and $180^{\circ}-\gamma_s\geq 90^{\circ}$) would appear even in the absence of the horizontal beam, in which case the perturbation reduces to a $1$D lattice with a periodicity that is symmetric over $\gamma=90^{\circ}$. The asymmetric structure (whose location we denote by $\gamma_a$) would appear in the absence of either of the oblique laser beams for similar reasons. The simulataneous appearance of all three structures requires the $2$D character of the lattice. Once the locations of these stability dips are measured, one can infer the roton wavelength via either $\lambda_{\mathrm{rot}}\approx \lambda_{\mathrm{Las}}/\left(2\sin\gamma_s\right)$ or $\lambda_{\mathrm{rot}}\approx\lambda_{\mathrm{Las}}/\left(2\cos\frac{\gamma_a}{2}\right)$. 

Similar to the $1$D-lattice scenario in Sec. \ref{sec: 1D latt}, mode-matching plays a minimal role when a zero-tilt rotonized dBEC is perturbed by a triangular lattice. This may be understood by considering matrix elements of the $\mathcal{A}$ matrix (defined below Eq. \eqref{eq: BdG pert}) between modes $\vec k'$ and $\vec k$. In this case, both $\left\langle \varphi_{\vec k'}| U_c^{(1)}|\varphi_{\vec k}\right\rangle$ and $\left\langle\varphi_{\vec k'}|X^{(1)}|\varphi_{\vec k}\right\rangle$ are proportional to a sum of the Kronecker delta functions 
\begin{equation}
\delta_{|k_x'-k_x|,k_{\mathrm{Las}}(1+\cos\gamma)}\delta_{|k_y'-k_y|,k_{\mathrm{Las}}\sin\gamma}
\end{equation}
and
\begin{equation}
\delta_{k_x,k_x'}\delta_{|k_y'-k_y|,2k_{\mathrm{Las}}\sin\gamma} ,
\end{equation}
where mode-matching can occur essentially via either term in Eq. \eqref{eq: Uc tri}. This is in direct analogy with Fig. \ref{fig: mode match} except that matched modes connect either vertically by the vector $\pm 2k_{\mathrm{Las}}\sin\gamma\hat y$ or diagonally by the vector $\pm k_{\mathrm{Las}}(1+\cos\gamma)\hat x \pm k_{\mathrm{Las}}\sin\gamma \hat y$, the norm of which is $2k_{\mathrm{Las}}\cos\frac{\gamma}{2}$. Rotons cannot mode-match if both of these vectors are too large in magnitude to form chords in a circle of diameter $2k_{\mathrm{rot}}$. Such is the case for angles in the vicinity of $\gamma=60^{\circ}$ when $k_{\mathrm{Las}}>2k_{\mathrm{rot}}/\sqrt{3}$ (equivalently, $\lambda_{\mathrm{Las}}<\sqrt{3}\lambda_{\mathrm{rot}}/2$), as follows from geometric principles.
This raises the stability boundary somewhat for such angles, since the destabilization of rotons is then a second-order effect; nevertheless, no new stability features emerge.

The question naturally arises as to how polarization tilt might affect the stability spectrum of a rotonized gas in a triangular lattice.  We have found that certain features in Fig. \ref{fig: stab tri} wash out with increasing tilt, depending both on $\lambda_{\mathrm{L}}/\lambda_{\mathrm{rot}}$ and the direction of polarization tilt. For the case of tilted dipoles, the amplification factors in Eq. \eqref{eq: A and B} are written more generally as
\begin{equation}
\begin{aligned}
& A_{k_{\mathrm{Las}},\gamma} = \frac{\left(2k_{\mathrm{Las}}\cos\left(\frac{\gamma}{2}\right)\right)^2}{2\varepsilon\left(\pm k_{\mathrm{Las}}(1+\cos\gamma)\hat x \pm k_{\mathrm{Las}}\sin\gamma \hat y   \right)} \\&
B_{k_{\mathrm{Las}},\gamma}= \frac{\left(2k_{\mathrm{Las}}\sin\gamma\right)^2}{2\varepsilon\left(\pm 2k_{\mathrm{Las}}\sin\gamma \hat y\right)}
\end{aligned}.
\end{equation}
Since the polarization tilt raises the relative Hartree-Fock energies of roton modes along the tilt projection, this results in $A_{k_{\mathrm{Las}},\gamma}$ and/or $B_{k_{\mathrm{Las}},\gamma}$ being less-strongly-peaked functions of $\gamma$. For example, tilting the polarization towards $\hat y$ raises the relative value of $\varepsilon (\pm k_{\mathrm{rot}}\hat y)$, which diminishes the enhancement due to $B_{k_{\mathrm{Las}},\gamma}$. This would cause a washing out of the symmetrically-located features in Fig. \ref{fig: stab tri}. Similarly, a polarization tilt towards $\hat x$ would diminish the enhancement due to $A_{k_{\mathrm{Las}},\gamma}$ and wash out the central stability feature.

We note that the stability structures originating from mean-field enhancement, both in the $1$D and $2$D lattices, can be observed in systems whose dispersions are not quite rotonized (ie, possess no roton minimum in $\omega(\vec k)$). This is because the Hartree-Fock energy $\varepsilon(\vec k)$, whose inverse appears in the combined potentials \eqref{eq: Uc 1d} and \eqref{eq: Uc tri}, may have a local minimum even when the dispersion $\omega(\vec k)$ does not exhibit a roton. In such cases, the inverse Hartree-Fock energy $1/\varepsilon(\vec k)$ is less strongly peaked, resulting in respective stability dips that are generally observable, albeit less pronounced. The locations of these dips are set by the local minimum of $\varepsilon(\vec k)$, and this is strictly true even for rotonized systems, in which case $k_{\mathrm{rot}}$ only approximates the location of the Hartree-Fock minimum. These particular structures thus do not measure the onset of rotonization in the dispersion, but rather the closely-related phenomenon of a Hartree-Fock local minimum at nonzero momentum. Such can occur at lower densities than are generally necessary for rotonization, making our results directly relevant to current dBEC experiments that do not appear to have fully rotonized dispersions.

\section{Conclusion} \label{sec: conclusion}

Stability spectroscopy is a promising new avenue for measuring the experimentally-elusive roton. When a rotonized dipolar condensate is perturbed by a weak optical lattice, the resulting stability plot contains spectroscopic information from which the roton wavelength may be inferred. In this paper, we have examined the stability spectra of rotonized q2D dipolar BECs in the presence of tunable $1$D and $2$D lattices. The tilt of the dipole polarization axis plays an important role in determining which structures are present in the $1$D-lattice stability diagram. The central stability feature at periodicity $\lambda_{\mathrm{L}}\approx \lambda_{\mathrm{rot}}$ exists independently of polarization tilt, whereas the features at $\lambda_{\mathrm{rot}}/2$ and $2\lambda_{\mathrm{rot}}$ emerge only when the rotonized gas has the polarization axis tilted into the plane, remaining orthogonal to the lattice vector. The emergent structures originate in the mode-matching of directional rotons, the relative energies of which depend on polarization tilt. In the case of a triangular lattice perturbing a zero-tilt dBEC, we find structure in the stability spectrum resulting from the $2$D character of the perturbation. The locations of stability dips can be tuned by varying the off-resonant laser wavelength used to produce the lattice. Recent advances in the condensation of magnetically dipolar atoms suggest that measurements of stability spectra should be feasible, and that rotons in dBECs may be consequently observed.

\section*{Acknowledgement}
J.P.C acknowledges support from the US DoD through the NDSEG fellowship program. R.M.W. acknowledges support from an NRC postdoctoral fellowship. J.L.B. acknowledges financial support from the NSF.


\begin{thebibliography}{99}

\bibitem{Leggett2001} A. J. Leggett, Rev. Mod. Phys. {\bf 73}, 307 (2001).



\bibitem{Griesmaier2005} A. Griesmaier, J. Werner, S. Hensler, J. Stuhler, and T. Pfau, Phys. Rev. Lett. {\bf 94}, 160401 (2005).

\bibitem{Aikawa2012} K. Aikawa, A. Frisch, M. Mark, S. Baier, A. Rietzler, R. Grimm, and F. Ferlaino, Phys. Rev. Lett. {\bf 108}, 210401 (2012).

\bibitem{Lu2011} M. Lu, N. Q. Burdick, S. H. Youn, and B. L. Lev, Phys. Rev. Lett. {\bf 107}, 190401 (2011).

\bibitem{Deiglmayr2008}J. Deiglmayr, A. Grochola, M. Repp, K. M\"{o}rtlbauer, C. Gl\"{u}ck, J. Lange, O. Dulieu, R. Wester, and M. Weidem\"{u}ller, Phys. Rev. Lett.{\bf 101}, 133004 (2008).

\bibitem{Ni2010}K.-K Ni, S. Ospelkaus, D. Wang, G. Qu\`{e}mener, B. Neyenhuis, M. H. G. de Miranda, J. L. Bohn, J. Ye, and D. S. Jin, Nature (London) {\bf 464}, 1324 (2010).

\bibitem{Ospelkaus2010}S. Ospelkaus, K.-K. Ni, G. Qu\`{e}mener, B. Neyenhuis, D. Wang, M. H. G. de Miranda, J. L. Bohn, J. Ye, and D. S. Jin, Phys. Rev. Lett. {\bf 104}, 030402 (2010).

\bibitem{Heidemann2008} R. Heidemann, U. Raitzsch, V. Bendkowsky, B. Butscher, R. L\"{o}w, and T. Pfau, Phys. Rev. Lett. {\bf 100}, 033601 (2008).

\bibitem{Koch2008} T. Koch, T. Lahaye, J. Metz, B. Fr\"{o}lich, A. Griesmaier, and T. Pfau, Nature Phys. {\bf 4}, 218 (2008).

\bibitem{Lahaye2008} T. Lahaye, J. Metz, B. Fr\"{o}hlich, T. Koch, M. Meister, A. Griesmaier, T. Pfau, H. Saito, Y. Kawaguchi, and M. Ueda, Phys. Rev. Lett. {\bf 101}, 080401 (2008).

\bibitem{Billy2012} J. Billy, E. A. L. Henn, S. M\"{u}ller, T. Maier, H. Kadau, A. Griesmaier, M. Jona-Lasinio, L. Santos, and T. Pfau, Phys. Rev. A {\bf 86}, 051603(R) (2012).

\bibitem{Santos2003} L. Santos, G. V. Shlyapnikov, and M. Lewenstein, Phys. Rev. Lett. {\bf 90}, (250403) (2003).

\bibitem{ODell2003} D. H. J. O'Dell, S. Giovanazzi, and G. Kurizki, Phys. Rev. Lett. {\bf 90}, 110402 (2003).

\bibitem{Matuszewski2010} Spin-orbit-coupled gases are also predicted to exibit rotons: M. Matuszewski, Phys. Rev. Lett. {\bf 105}, 020405 (2010).

\bibitem{Henshaw1958} D. G. Henshaw, Phys. Rev. Lett. {\bf 1}, 127 (1958).

\bibitem{Palevsky1958} H. Palevsky, K. Otnes, and K. E. Larsson, Phys. Rev. {\bf 112}, 11 (1958).

\bibitem{Yarnell1959} J. L. Yarnell, G. P. Arnold, P. J. Bendt, and E. C. Kerr, Phys. Rev. {\bf 113}, 1397 (1959).

\bibitem{Henshaw1961} D. G. Henshaw and A. D. B. Woods, Phys. Rev. {\bf 121}, 1266 (1961).

\bibitem{Fischer2006} U. R. Fischer, Phys. Rev. A {\bf 73}, 031602(R) (2006).

\bibitem{Boudjemaa2012} A. Boudjem\^{a}a and G. V. Shlyapnikov, Phys. Rev. A {\bf 87}, 025601 (2013).



\bibitem{Faraday} R. Nath and L. Santos, Phys. Rev. A {\bf 81}, 033626 (2010); K. Lakomy, R. Nath, and L. Santos, Phys. Rev. A {\bf 86}, 023620 (2012).

\bibitem{Sykes2012} A. G. Sykes and C. Ticknor, arXiv:1206:1350 (2012).

\bibitem{Santos2013} M. Jona-Lasinio, K. Lakomy, and L. Santos, arXiv:1301.4907 (2013).

\bibitem{Wilson2012} R. M. Wilson, C. Ticknor, J. L. Bohn, and E. Timmermans, Phys. Rev. A {\bf 86}, 033606 (2012).

\bibitem{Mottl2012} R. Mottl, F. Brennecke, K. Baumann, R. Landig, T. Donner, and T. Esslinger, Science {\bf 336}, 1570 (2012).

\bibitem{Blakie2012} P. B. Blakie, D. Baillie, and R. N. Bisset, Phys. Rev. A {\bf 86}, 021604(R) (2012).

\bibitem{Corson2013} J. P. Corson, R. M. Wilson, and J. L. Bohn, Phys. Rev. A (R) (In Press).

\bibitem{Grimm2000} R. Grimm, M. Weidem\"{u}ller, and Y. Ovchinnikov, Adv. At. Mol. Opt. Phys. {\bf 42}, 95 (2000).

\bibitem{Yi2000} S. Yi and L. You, Phys. Rev. A {\bf 61}, 041604(R) (2000).

\bibitem{Petrov2000} D. S. Petrov, M. Holzmann, and G. V. Shlyapnikov, Phys. Rev. Lett. {\bf 84}, 2551 (2000).

\bibitem{Ticknor2011} C. Ticknor, R. M. Wilson, and J. L. Bohn, Phys. Rev. Lett. {\bf 106}, 065301 (2011).

\bibitem{Wilson2009} R. M. Wilson, S. Ronen, and J. L. Bohn, Phys. Rev. A {\bf 80}, 023614 (2009).

\bibitem{Fetter} A. Fetter and J. Walecka, {\it Quantum Theory of Many-Particle Systems}, (Dover, New York, 1971).

\bibitem{Ticknor2012} C. Ticknor, Phys. Rev. A {\bf 86}, 053602 (2012).

\bibitem{Wilson2008} R. M. Wilson, S. Ronen, J. L. Bohn, and H. Pu, Phys. Rev. Lett. {\bf 100}, 245302 (2008).

\bibitem{Taylor2003} E. Taylor and E. Zaremba, Phys. Rev. A {\bf 68}, 053611 (2003).

\bibitem{Gaul2011} C. Gaul and C. A. M\"{u}ller, Phys. Rev. A {\bf 83}, 063629 (2011).

\bibitem{Liang2008} Z. X. Liang, X. Dong, Z. D. Zhang, and B. Wu, Phys. Rev. A {\bf 78}, 023622 (2008).

\bibitem{Lugan2011} P. Lugan and L. Sanchez-Palencia, Phys. Rev. A {\bf 84}, 013612 (2011).

\bibitem{Pitaevskii} L. Pitaevskii and S. Stringari, {\it Bose-Einstein Condensation}, (Clarendon Press, Oxford, 2003).

\bibitem{Press} W. Press. S. Teukolsky, W. Vetterling, and B. Flannery, {\it Numerical Recipes in FORTRAN}, $2^{\mathrm{nd}}$ {\it Ed}. (Cambridge University Press, New York, 1992).

\bibitem{Kuhn2013} S. K\"{u}hn and T. E. Judd, Phys. Rev. A {\bf 87}, 023608 (2013).

\bibitem{Morsch2006} O. Morsch and M. Oberthaler, Rev. Mod. Phys. {\bf 78}, 179 (2006).

\bibitem{symmetry} The experiment could also be performed by keeping two beams stationary while varying the direction of a third intersecting beam. Such a scenario is experimentally simpler, and the locations of stability dips can still be determined by considering the mean-field enhancement of the combined potential, as in Eq. \eqref{eq: Uc tri}. 

\bibitem{Coordinates} We have ignored the three relative phases of the lasers, as they do not affect the mode matching of rotons to the lattice. 



\bibitem{Pethick} C. J. Pethick and H. Smith, {\it Bose-Einstein Condensation in Dilute Gases}, $2^{\mathrm{nd}}$ {\it Ed}. (Cambridge University Press, Cambridge, 2008).


\end{thebibliography}

\end{document}